\documentclass{ws-ijmpcs}
\usepackage{booktabs}

\usepackage{slashed}
\usepackage{amssymb,amsmath}

\usepackage{ulem}
\usepackage{color}

\newcommand{\Eqref}[1]{Eq.~\eqref{#1}}
\newcommand{\Tmin}{T_{\text{min}}}

\begin{document}

\markboth{Sch\"{a}fer, Huet, Gies}
{Casimir Energy-Momentum Tensors \\ with Worldline Numerics}

%
\catchline{}{}{}{}{}
%

\title{ENERGY-MOMENTUM TENSORS WITH WORLDLINE NUMERICS}

\author{MARCO SCH\"{A}FER$^1$, IDRISH HUET$^{1,2}$, HOLGER GIES$^{1,2}$}

\address{$^1$ Theoretisch-Physikalisches Institut, Friedrich-Schiller-Universit\"{a}t Jena\\
Max-Wien-Platz 1, D-07743 Jena, Germany\\
$^2$ Helmholtz Institute Jena, Fr\"{o}belstieg 3, D-07743 Jena, Germany\\
marco.schaefer@uni-jena.de, idrish.huet@uni-jena.de, holger.gies@uni-jena.de}

\maketitle

\begin{history}
\received{Day Month Year}
\revised{Day Month Year}
\end{history}

\begin{abstract}
  We apply the worldline formalism and its numerical Monte-Carlo approach to
  computations of fluctuation induced energy-momentum tensors. For the case of
  a fluctuating Dirichlet scalar, we derive explicit worldline
    expressions for the components of the canonical energy-momentum tensor
    that are straightforwardly accessible to partly analytical and generally
    numerical evaluation. We present several simple proof-of-principle
    examples, demonstrating that efficient numerical evaluation is possible at
  low cost. Our methods can be applied to an investigation of positive-energy conditions.


\end{abstract}

\section{Introduction}

It has been over five decades since Casimir predicted a negative
{interaction} energy {density} and a
corresponding attractive force\cite{Casimir:dh} between two perfectly conducting plates in a
vacuum which impose boundary conditions on the quantized electromagnetic
field. Although the theory of the Casimir effect has
  become a well developed field in recent years\cite{Bordag:2001qi}, it
  still serves as a paradigmatic test case for exploring the diversity of
  phenomena within quantum field theories. In the first place, interaction energies, pressures
and forces caused by boundary conditions (BCs) imposed on a quantum
field are of particular 
  relevance, as they constitute phenomenological observables of the Casimir
  effect and can be directly related to experiment. From a more theoretical perspective, this information is more
generally
  encoded in the expectation value of the
energy-momentum tensor (EMT) of the fluctuating field. 

Beyond the direct phenomenological application, the
  energy-momentum tensor is also at the center of interest since it serves as a source term for the gravitational
field if the quantum field is coupled to gravity. For instance, the energy-momentum tensor is needed in order to
  study how the Casimir energy falls in a gravitational field\cite{hep-th/0702091}. For usual Casimir force calculations, the
  fact that Casimir energies can generically be negative is not a
  problem at all, as it is similar to negative binding energies in
  bound state systems. However, in connection with general relativity,
  negative energy densities can be puzzling. Such energy densities may
  be in conflict with restrictions imposed on the energy-momentum
  tensor in order to avoid exotic phenomena such as superluminal
  travel, wormholes or time machines. Indeed, the Casimir effect
  violates particular restrictions such as the weak energy condition
  (WEC), $\langle \hat T_{\mu\nu} V^\mu V^\nu \rangle\geq 0$ with
  $V^\mu$ being a timelike vector, or the null energy condition (NEC),
  $\langle \hat T_{\mu\nu} V^\mu V^\nu \rangle\geq 0$ with $V^\mu$
  being a null vector\cite{gr-qc/0205134,hep-th/0307067}. By
  contrast, both conditions are obeyed by 
  classical physics. 

A somewhat weaker condition that is still sufficient to rule out
  exotic phenomena is the averaged null energy condition (ANEC) which requires that the null
  energy condition has to hold only when integrated along a complete geodesic. The ANEC has been found to
be satisfied in all Casimir examples studied so far on flat Minkowski space (even
  though it can be violated on compact flat or curved spaces\cite{gr-qc/9604008}). An investigation of the ANEC in specific Casimir configurations is
 typically involved, as certain components of the energy-momentum
 tensor need to be known along a complete geodesic. However, the
 physically interesting geodesics which appear to
 collect negative energy densities along the geodesic are typically
 directed towards a surface and thus hit the surface at some point. In
 such a case, the average acquires a generically large positive
 contribution from the surface itself which -- though nonuniversal --
 is likely to exceed the negative Casimir contributions by far. Relevant
 configurations that avoid the discussion of large nonuniversal contributions therefore need to have
holes through which a geodesic can
 pass. In fact, the case of a single plate with a hole has been found
 to obey the ANEC\cite{hep-th/0506136}.

As is obvious from these considerations, a general study of
  energy conditions in Casimir configurations requires a theoretical
  framework that is capable of dealing with the energy-momentum tensor
  in arbitrary geometries. Standard computations of Casimir energy
  momentum tensors\cite{55030,hep-th/0210081,quant-ph/0507042,quant-ph/0408163,arXiv:0911.2688}
  are usually based on mode summation/expansion, image charge methods
  or similar techniques which are suited for specific simple
  geometries. In the present work, we report on progress using the
  worldline approach\cite{Schubert:2001he} to the Casimir effect for the case of a Dirichlet
  scalar field\cite{Gies:2003cv,Gies:2006cq} which has meanwhile been
  used to study a variety of nontrivial Casimir configurations\cite{Gies:2006xe,Weber:2009dp,814203,arXiv:1110.5936}. We show that
  the worldline approach can straightforwardly be generalized to study
  Casimir energy-momentum tensors. We discuss the similarities and
  differences with conventional interaction energy computations and
  verify the resulting algorithm with the help of known simple
  examples.

\section{Worldline Formalism for composite operators}

We investigate the energy-momentum tensor of a 
minimally coupled quantum
scalar field $\hat{\Phi}$ with zero mass which shall be defined on a
$d+1$-dimensional domain $\mathcal D$. On the boundary
$\partial\mathcal D$ that we consider as static
we impose Dirichlet boundary conditions. The canonical energy-momentum
tensor operator of $\hat{\Phi}$ reads
\begin{equation}
\hat{T}_{\mu\nu}(x,t):=\lim_{{x}\to {x}^\prime}
\left[\partial_\mu\hat{\Phi}\partial^\prime_{\nu}\hat{\Phi}^\prime-\frac{1}{2}g_{\mu\nu}
\left(
\partial_\alpha\hat{\Phi}\partial^{\prime\alpha}\hat{\Phi}^\prime-\sigma(x)\hat{\Phi}\hat{\Phi}^\prime\right)\right], \label{eq:1}
\end{equation}
where we have employed a point-splitting procedure in the spatial
coordinates for regularization. 
The background potential $\sigma(x)$ 
can in principle be arbitrary but will later specifically be used to impose Dirichlet BCs. For 
our study of the energy conditions, we need to compute the vacuum expectation value
$\left\langle\hat{T}_{\mu\nu}\right\rangle$ of the EMT. For instance, for the null energy condition, we have
to evaluate the combination
\begin{equation} \label{TVV}
 \left\langle\hat{T}_{\mu\nu}V^{\mu}V^{\nu}\right\rangle=\left\langle\hat{T}_{00}+\hat{T}_{zz}\right\rangle=T_{00}+T_{zz},
\end{equation}
where we have choosen the null vector $V^{\mu}$ in $z$ direction: $V^{\mu}=(1,0,\ldots,0,1)$. 
In a next step we expand the scalar field $\hat{\Phi}$ in terms of momentum modes
$\psi_n$. These momentum modes are defined as eigenmodes of the Laplacian in the presence of $\sigma(x)$, 
\begin{equation}
 \left(-{\nabla}^2+\sigma(x)\right)\psi_n(x)=k_n^2 \psi_n(x),
\label{eq:Laplace}
\end{equation}
with $k_n^2$ denoting the eigenvalue. We now introduce the propagator
$G({x},{x}^\prime,k)$ of the corresponding Helmholtz equation
\begin{equation}
{ \left(-{\nabla}^2+\sigma(x)-
    k^2\right)G({x},{x}^\prime,k)=\delta({x}-{x}')}
\label{eq:Geq}
\end{equation}
which has the spectral representation
\begin{equation}
G({x},{x}^\prime,k)= \sum_n \frac{\psi_n(x)\psi_n^{*}(x^\prime)}{
  k_n^2-k^2-i\varepsilon}.
\end{equation}
The components of the energy-momentum tensor in Eq. (\ref{TVV}) can now be
expressed solely in terms of $G({x},{x}^\prime,k)$. As we are
interested in the effect of the background potential $\sigma(x)$ in comparison to the vacuum case $\sigma=0$,
we normalize the EMT with respect to the trivial vacuum by subtracting the free Green's function
$G_0({x},{x}^\prime,k)$ obtained from \Eqref{eq:Geq} for $\sigma=0$. This corresponds to removing the
$\sigma$-independent divergence,
\begin{align} \label{EMT00}
\langle \hat{T}_{00}({x}, t) \rangle = {} & \lim_{{x}\to{x}^\prime} \int_0^{\infty} \frac{\mathrm{d} k}{\pi}\left( k^2+\frac{1}{2}{\nabla}\cdot\left({\nabla}+{\nabla}^\prime\right) \right)\operatorname{Im}\left[(G-G_0)({x},{x}^\prime,k)\right],\\ \label{EMTzz}
	\langle \hat{T}_{zz}({x}, t) \rangle = {} & \lim_{{x}\to{x}^\prime} \int_0^{\infty} \frac{\mathrm{d} k}{\pi}\left( \partial_z\partial_{z^\prime}-\frac{1}{2}{\nabla}\cdot\left({\nabla}+{\nabla}^\prime\right) \right)\operatorname{Im}\left[(G-G_0)({x},{x}^\prime,k)\right].
\end{align}
The whole formalism works, of course, for any field configuration $\sigma(x)$. In the following, we focus on
$\sigma$ backgrounds that induce Dirichlet BCs on $\partial\mathcal{D}$. For this, we schematically use
\begin{equation}
 \sigma(x)=\lambda\,\delta\left({x-x_{\partial\mathcal{D}}}\right),
 \quad {x_{\partial\mathcal{D}}\in \partial\mathcal{D},}
\end{equation}
and take the Dirichlet limit $\lambda\to\infty$ \cite{hep-th/0309130}. The EMT in Eq. (\ref{EMT00})    
and (\ref{EMTzz}) is then finite on $\mathcal{D}$ where the potential $\sigma$
vanishes. There remain, however, divergences on the boundary
$\partial\mathcal{D}$ due to the Dirichlet limit. These can be
  interpreted as the infinite amount of energy needed to constrain a scalar
  quantum field on all momentum scales to satisfy the Dirichlet constraint on
  the boundary\cite{hep-th/0309130}.

Interpreting the eigenvalue problem \eqref{eq:Laplace} as a ficticious
  quantum mechanical Schr\"{o}dinger problem with Hamiltonian $H=-\nabla^2 +
  \sigma(x)$, the Green's function corresponds to a quantum mechanical
  propagator Fourier transformed to energy space, which can be written as a
  Feynman path integral in position space. The vacuum-subtracted path integral
  representation of the propagator then results in
\begin{align}\label{GreenDiff}
(G-G_0)({x},{x}^\prime,k)
 = {} &
i\int\limits_0^\infty\mathrm{d}s\,e^{isk^2}\int\limits_{x^\prime{=x(0)}}^{x{=x(T)}}\mathcal{D}
{x}(\tau)e^{
i\int\limits_0^s\mathrm{d}\tau\frac{\dot{{x}}^2}{4}}\left(e^{-i\int\limits_0^s\mathrm{d}\tau\,
\sigma({x})}-1\right)\nonumber\\
= {} &
\int\limits_0^\infty\mathrm{d}T\,e^{-Tk_E^2}\int\limits_{x^\prime}^{x}\mathcal{D}{x}(\tau)e^{ 
-\int\limits_0^T\mathrm{d}\tau\frac{\dot{{x}}^2}{4}}\left(e^{-\int\limits_0^T\mathrm{d}\tau\,
\sigma({x})}-1\right),
\end{align}
where $s$ is a ficticious quantum mechanical time, so-called propertime. We have formally made Wick rotations
both in the $s$ and $k$ planes so that $s=-iT$ and $k_E = ik$ to cast the expression above in Euclidean form. 

In contrast to previous calculations of effective interaction energies for the Casimir effect
and similar setups which required path integrals only over closed
loops, here we need to compute path integrals of open worldlines from
${x}^\prime$ to ${x}$. This can conveniently be done with the
$d$ loop algorithm\cite{Gies:2005sb}.
The path integral is normalized here implicitly such that the free path
  integral yields the standard free propagator,
\begin{equation}
\int\limits_{x^\prime}^{x}\mathcal{D}x(\tau)e^{
-\int\limits_0^T\mathrm{d}\tau\frac{\dot{x}^2}{4}}
= \frac{e^{-\frac{({x}-{x}^\prime)^2}{4T}}} {\left(4\pi
    T\right)^{\frac{d}{2}}}. 
\label{eq:normal}
\end{equation}
This normalization is made explicit by the following shorthand notation
\begin{equation}
\frac{e^{-\frac{({x}-{x}^\prime)^2}{4T}}} {\left(4\pi
      T\right)^{\frac{d}{2}}}
\left\langle\mathcal{O}\right\rangle_{xx'}:=\int\limits_{x^\prime}^{x}\mathcal{D}x(\tau)e^{ 
-\int\limits_0^T\mathrm{d}\tau\frac{\dot{x}^2}{4}}\mathcal{O}.
\end{equation}
Then Eq. (\ref{GreenDiff}) can be written as 
\begin{align}\label{GreenDiffExpVal}
 (G-G_0)({x},{x}^\prime,k=-ik_E) = {} & \int\limits_0^\infty\mathrm{d}T\,e^{-Tk_E^2}
\frac{e^{-\frac{({x}-{x}^\prime)^2}{4T}}}{\left(4\pi
  T\right)^{\frac{d}{2}}} \left\langle
e^{-\int\limits_0^T\mathrm{d}\tau\,\sigma(x)}-1\right\rangle_{{xx'}} . 
\end{align}
In the Dirichlet limit the last term can be simplified
\begin{equation}
  \left\langle
  e^{-\int\limits_{0}^T\mathrm{d}\tau\,\sigma(x)}-1\right\rangle_{{xx'}}
  \!\!\!\!\!
=
\left\{\begin{array}{cl} -1 & \mbox{if $x(\tau)$ intersects a boundary}\\ 0 & \mbox{{otherwise}} \end{array}
\right\}=-\left\langle\Theta\left(T-\Tmin\right)\right\rangle_{{xx'}}
\end{equation}
which means that only loops $x(\tau)$ that violate the boundary
  conditions lead to deviations from the trivial vacuum and thus
contribute to the path integral. The quantity $\Tmin$ denotes the
  minimum propertime needed for a given worldline to propagate from $x'$ to
  $x$ and simultaneously intersect a boundary in between. If $T$ is too small
  the particle has to propagate on a straight line from $x'$ to
  $x$. If $T$ is sufficiently long the diffusive Brownian motion process behind
  the path integral creates random detours that can eventually intersect the
  boundaries. We may choose our integration contour in the $k_E$ plane in such a manner that the imaginary part of
the integral is already taken into account by carrying out the $k_E$ 
integrations in (\ref{EMT00}) and (\ref{EMTzz}) along the real axis, indeed this is the effect of the Wick
rotation. These considerations lead us to {a compact representation
  of the EMT components required, e.g., for the null energy condition: 
\begin{align}\label{T00one}
\langle \hat{T}_{00}({x}, t) \rangle(I) = {} &  
\frac{1}{(4\pi)^{\frac{d+1}{2}}}\int\limits_0^\infty\frac{\mathrm{d}T}{2T^{\frac{d+3}{2}}}
e^{-\frac{({x}-{x}^\prime)^2}{4T}}\left\langle\Theta\left(T-\Tmin\right)\right\rangle_{{xx'}},\\  
\label{Tzzone}\langle \hat{T}_{zz}({x}, t) \rangle(I) = {} &  
-\frac{1}{(4\pi)^{\frac{d+1}{2}}}\partial_{z}\partial_{z^{\prime}}
\int\limits_0^\infty\frac{\mathrm{d}T}{T^{\frac{d+1}{2}}}e^{-\frac{({x}-{x}^\prime)^2}{4T}}
\left\langle\Theta\left(T-\Tmin\right)\right\rangle_{{xx'}},\\
\langle \hat{T}_{00}({x}, t) \rangle(II) = {} &
-\frac{1}{(4\pi)^{\frac{d+1}{2}}}{\nabla}\cdot\left({\nabla}+{\nabla}
^\prime\right)\int\limits_0^\infty\frac{\mathrm{d}T}{2T^{\frac{d+1}{2}}}
e^{-\frac{({x}-{x}^\prime)^2}{4T}}\left\langle \Theta\left(T-\Tmin\right)\right\rangle_{{xx'}}\nonumber\\
= {} & -\langle \hat{T}_{zz}({x}, t) \rangle(II), \label{eq:T00two}
\end{align}
where the limit ${x}\to{x}^\prime$ is implicitly understood from now on. From
  the last identity, it is, for instance, obvious that the NEC only involves
$\left\langle\hat{T}_{\mu\nu}V^{\mu}V^{\nu}\right\rangle=T_{00}(I)+T_{zz}(I)$. We emphasize that the angle brackets
are used in two different contexts: on the l.h.s. $\left\langle\dots\right\rangle$ denotes the vacuum
expectation of a composite quantum operator, whereas on the r.h.s. $\left\langle\dots\right\rangle_{{xx'}}$
denotes the expectation value of the path integral, that is the average over
an ensemble of open worldlines.

\section{Energy momentum tensor for a single plate}

With these general considerations, we can now compute
the EMT components in the case where the boundary is one single plate at $z=0$
with the $z$ axis being its normal.

\subsection{Analytical calculation for the single plate}

Analytical results for the single plate are easily obtained through explicit calculation
of the Green's function using the method of images. After regularization one finds
\begin{equation}
(G-G_0)({x},{x}';k) = - \varphi({\xi}, {x}')
\end{equation}
in which $\varphi({\xi},{x}')$ denotes the field at
${x}':=({x}'_{\parallel},z')$ produced by a point source sitting at
${\xi}:=({x}_{\parallel},-z)$. Also ${x}:=({x}_{\parallel},z)$, with
$z$ and $z'$ being the distances from the plate. The
explicit forms of the field $\varphi$ in two and three dimensions are
given by $\varphi_{d=2} = \frac{1}{2\pi} K_{0} (-i k|{x}-{x}'|)$ and
$\varphi_{d=3} = \frac{ e^{i k|{x} -{x}'|}}{4\pi |{x} - {x}'|}$,
respectively. Using them it is straightforward to show that
\begin{align}
\left\langle\hat{T}_{00}\right\rangle_{\mathrm{d=2}}\mathrm{(I)}\  &= \quad\!\frac{1}{32\pi}
\frac{1}{z^3},
&\left\langle\hat{T}_{00}\right\rangle_{\mathrm{d=3}}\mathrm{(I)}\  &= \quad\!\frac{1}{32\pi^2}\frac{1}{z^4},\label{eq:SPa}\\
\left\langle\hat{T}_{00}\right\rangle_{\mathrm{d=2}}\mathrm{(II)} &=  -\frac{1}{16
\pi}\frac{1}{z^3}, 
&\left\langle\hat{T}_{00}\right\rangle_{\mathrm{d=3}}\mathrm{(II)} &=  -\frac{3}{32
\pi^2}\frac{1}{z^4},\\ 
\left\langle\hat{T}_{zz}\right\rangle_{\mathrm{d=2}}\mathrm{(I)}\  &=  -\frac{1}{16
\pi}\frac{1}{z^3}, 
&\left\langle\hat{T}_{zz}\right\rangle_{\mathrm{d=3}}\mathrm{(I)}\  &= -\frac{3}{32
\pi^2}\frac{1}{z^4}. 
\end{align}

\subsection{Worldline calculation for the single plate}

For the computation of the expectation value in Eq. (\ref{T00one}) and
(\ref{eq:T00two}), we need to evaluate the
{worldline} $\Theta$ function, i.e., we need to find the condition
under which a loop in our ensemble intersects the plate (see 
Fig.~\ref{Pic1PlateSetup} for a sketch).
\begin{figure}[ht]
\centerline{\includegraphics[width=5cm]{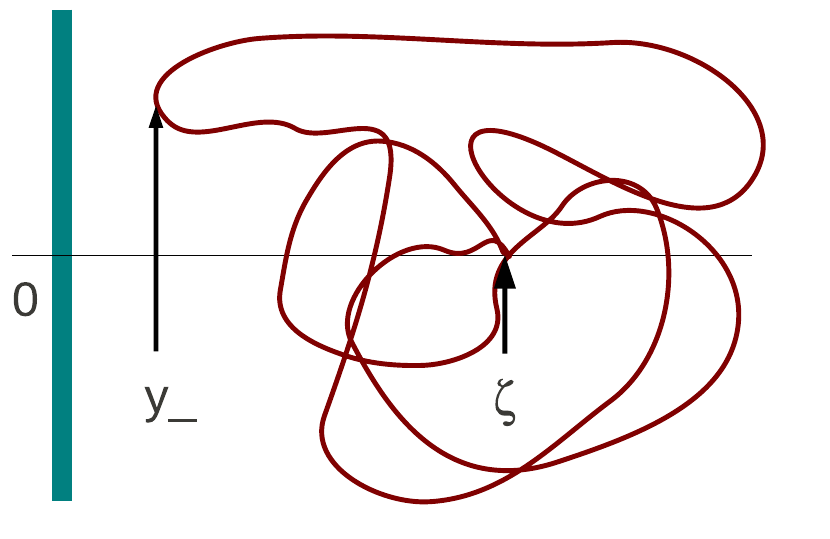}}
\caption{Sketch of the single plate setup. \label{Pic1PlateSetup}}
\end{figure}

After rescaling the loops for numerical convenience\cite{Gies:2003cv} 
\begin{equation}
x(Tt)\to\sqrt{T}y(t),\,\,t\in [0,1]\quad \Rightarrow\quad
\int_{x'}^{x} \mathcal{D}x(\tau) e^{ 
-\int\limits_0^T\mathrm{d}\tau\frac{\dot{x}^2}{4}} \to
\int_{x'/\sqrt{T}}^{x/\sqrt{T}} \mathcal{D}y(t) e^{ 
-\int\limits_0^1\mathrm{d}t\frac{\dot{y}^2}{4}}, 
\end{equation}
we find in the limit $x'\to x$, that is, $z'\to z$, the intersection condition
\begin{equation}
\sqrt{T}y_{-}+z\leq 0\  \Longrightarrow\
\Theta\left(T-T_{\text{min}}\right)=\Theta\left(T-\frac{z^2}{y_{-}^2}\right),
\end{equation}
where $z$ denotes the distance from the plate and $y_{-}$ denotes the
point of a loop that is closest to the plate, see Fig.~\ref{Pic1PlateSetup}.
The quantity $y_{-}$ is the only information we need about each loop. Since the worldline
  distributions factorize with respect to their position space components, we only need 1-dimensional loops. In
$d=2$ the $T_{00}$ component of the EMT then reads
\begin{align}
\left\langle\hat{T}_{00}\right\rangle=\left\langle\hat{T}_{00}\right\rangle(\text{I})+
\left\langle\hat{T}_{00}\right\rangle(\text{II}) = {} & \frac{1}{(4\pi)^{\frac{3}{2}}}\frac{1}{z^3}\left(
\frac{\langle \left|y_{-}\right|^3\rangle}{3}-\langle\left|y_{-}\right|\rangle\right)\\
 = {} & \frac{1}{(4\pi)^{\frac{3}{2}}}\frac{1}{z^3}\left(\frac{\sqrt{\pi}}{4}-\frac{\sqrt{\pi}}{2}
\right)
\end{align}
with the corresponding analytical values in the second line. In Table~\ref{Tab1Pl2d}
we compare our analytical and numerical 
results for two different ensembles with $N$ loops per ensemble and
$ppl$ points per loop.
\begin{table}[t]
\tbl{Worldline results for the single plate in $d=2$.}{
\begin{tabular}{@{}ccccc@{}}
\toprule
$ppl$, $N$ & $\langle\left|y_{-}\right|\rangle$ & $\frac{\sqrt{\pi}}{2}$ 
& $\langle\left|y_{-}\right|^{3}\rangle$ & $\frac{3\sqrt{\pi}}{4}$ \\ 
\colrule
$2^{14},\, 10^{4}$ & $0.8789\pm 0.0046$ & 0.8862 & $1.3018\pm 0.0210$ & 1.3293 \\
\midrule
$2^{18},\, 5\cdot10^{5}$ & $0.8841\pm 0.0007$ & 0.8862 & $1.3236\pm 0.0029$ & 1.3293 \\
\botrule
\end{tabular}\label{Tab1Pl2d}}
\end{table}
The error corresponds to the statistical uncertainty only and it
decreases like $1/\sqrt{N}$ as expected. 
The systematic error that is not displayed explicitly is governed by
the discretization of the loops (i.e. by $ppl$). A 
larger number of points per loop improves our results. As is known
from previous calculations of the Casimir energy 
using worldline numerics the numerical values approach the analytical
values from below, that is the discretized loops 
are smaller than the continuous ones.

The same analysis can be done for three spatial dimensions where we find
\begin{align}
\left\langle\hat{T}_{00}\right\rangle = {} &
\frac{1}{(4\pi)^{2}}\frac{1}{z^4}\left( 
\frac{\langle
  \left|y_{-}\right|^4\rangle}{4}-\frac{3\langle\left|y_{-}\right|^2\rangle}{2}\right)\\ 
 = {} & \frac{1}{(4\pi)^{2}}\frac{1}{z^4}\left(\frac{1}{2}-\frac{3}{2} 
\right).
\end{align}
The structure of the result is, of course, identical but the exponents of our loop variable and coefficients have
changed. In Table~\ref{Tab1Pl3d} we show our worldline results. Just as in the two-dimensional case we again see
the decrease of errors with increasing number of loops and
$ppl$. In both cases, these results confirm the violation
  of the weak energy condition by the single-plate configuration in
  agreement with Ref.~\refcite{gr-qc/0205134}.
\begin{table}[ht]
\tbl{Worldline results for the single plate in $d=3$.}{
\begin{tabular}{@{}ccccc@{}}
\toprule
$ppl$, $N$ & $\langle\left|y_{-}\right|^2\rangle$ & $1$ & $\langle\left|y_{-}\right|^{4}\rangle$ & $2$\\
\colrule
$2^{14},\, 10^{4}$ & $0.9841\pm 0.0100$ & $1.0000$ & $1.9610\pm 0.0488$ & $2.0000$\\
\midrule
$2^{18},\, 5\cdot10^{5}$ & $0.9964\pm 0.0014$ & $1.0000$ & $1.9901\pm 0.0063$ & $2.0000$\\
\botrule
\end{tabular}\label{Tab1Pl3d}}
\end{table}

As a further check of our algorithm we compute the null energy
condition along the $z$ axis 
\begin{align} \label{nullcondition}
  \left\langle\hat{T}_{00}+\hat{T}_{zz}\right\rangle = {} & \frac{1}{(4\pi
  )^{\frac{d+1}{2}}}\frac{1}{z^{d+1}}\left(\frac{\langle\left|y_{-}\right|^{d+1}\rangle} 
  {d+1}-\frac{d}{2}\langle\left|y_{-}\right|^{d-1}\rangle\right)
\end{align}
which is also violated\cite{gr-qc/0205134}. Comparing with the
analytical results again shows that our algorithm 
reproduces the expected values within the numerical precision set by
the parameters of our loop 
ensemble, viz. the number of points per loop $ppl$ and the number of loops $N$.

\section{Energy density for two parallel plates}

Let us now turn to the classic Casimir configuration of two
  infinite parallel plates separated by a distance $a$ imposing
  Dirichlet BCs on the fluctuating scalar field. However, because our
energy-momentum tensor is finite everywhere except for
$\partial\mathcal{D}$ we are not only able to compute energy
conditions or EMT components but actually also parts of components. In
the following, we focus on $T_{00}(I)$, cf. \Eqref{T00one}. 

\subsection{Analytical calculation for parallel plates}

As before $T_{00}(I)$ may be calculated by means of the traditional Green's function approach and the method of
images which gives
\begin{equation}
(G-G_0) ({x},{x}';k) = \sum_{q \in \mathbb{Z}/\{0\}}  \varphi({x}_q,{x}') - \sum_{ q \in
\mathbb{Z}} \varphi({\xi}_q, {x}').
\end{equation}
Here the vectors ${x}_q := ({x}_{\parallel}, z + 2qa),~
{\xi}_q : = ({x}_{\parallel}, -z+(2q+1)a)$ were
introduced. This time, $z$ measures the distance from the plates in a
coordinate system centered in the middle of the plates. {Introducing
the dimensionless coordinate $\zeta=z/a$} and inserting
the expressions for $\varphi$ given above \Eqref{eq:SPa} yields for
the region between the
plates where $|\zeta|\leq 
1/2$
\begin{align}
\left\langle\hat{T}_{00}\right\rangle_{\mathrm{d=2}}\mathrm{(I)} = {} & -\frac{1}{32\pi}\frac{1}{a^3}
\left(2\zeta_R\left(3\right)-\zeta_H\left(3,\frac{1}{2}+\zeta\right)-\zeta_H\left(3,\frac{1}{2}-
\zeta\right)\right),\nonumber\\
\left\langle\hat{T}_{00}\right\rangle_{\mathrm{d=3}}\mathrm{(I)} = {} & -\frac{1}{32\pi^2}\frac{1}{a^4}
\left(2\zeta_R\left(4\right)-\zeta_H\left(4,\frac{1}{2}+\zeta\right)-\zeta_H\left(4,\frac{1}{2}-
\zeta\right)\right), \label{CasimirT00I}
\end{align}
with $\zeta_{R,H}$ denoting the Riemann and Hurwitz zeta functions,
respectively.

\subsection{Worldline calculation for parallel plates}

The numerical worldline computation of
\begin{align}
\left\langle\hat{T}_{00}\right\rangle\mathrm{(I)} = {} &   
\frac{1}{(4\pi)^{\frac{d+1}{2}}}\int\limits_0^\infty\frac{\mathrm{d}T}{2T^{\frac{d+3}{2}}}
\left\langle 
\Theta\left(T-T_{\text{min}}\right)\right\rangle
\end{align}
for the Casimir plates is similar to the single-plate case. However,
as there are now two plates we need two 
intersection conditions as well. Accordingly we must compute two
points of each loop, viz. the points 
$y_{\pm}$ which are closest to the plates, see Fig.~\ref{PicCasimirSetup}.
\begin{figure}[ht]
\centerline{\includegraphics[width=5cm]{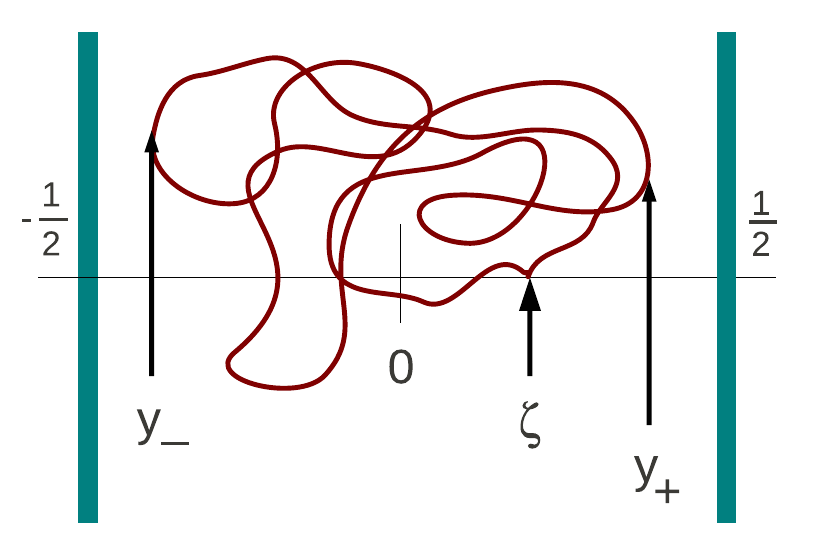}}
\caption{Sketch of Casimir's parallel-plates configuration in units of
  $a=1$. \label{PicCasimirSetup}}
\end{figure}
The corresponding value of $T_{\text{min}}$ is then easily found to be
\begin{equation}
\sqrt{T}y_{\pm}+{a}\left(\zeta\mp\frac{1}{2}\right)\leq 0\  \Longrightarrow\ 
\Theta\left(T-T_{\text{min}}\right)=\Theta\left(T- {a^2}
\min\left[\left(\frac{\zeta\pm\frac{1}{2}}{y_{\mp}} 
\right)^2\right ] \right).
\end{equation}
The analytical findings of \Eqref{CasimirT00I} compare favorably with
the numerical data displayed in Fig.~\ref{FCasimir}. 
\begin{figure}[ht]
\centerline{\includegraphics[width=7.7cm]{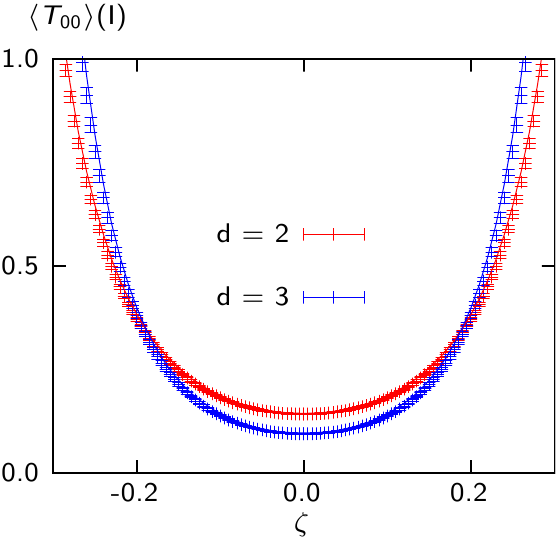}}
\caption{EMT component $T_{00}(I)$ in between two Casimir plates at a distance $a$
  in two and three spatial dimensions, $\zeta=z/a$. The analytical
  result \Eqref{CasimirT00I} (solid lines) is very well approximated by the
  numerical data.  \label{FCasimir}}
\end{figure}

\subsection{Conclusion}

We have studied Casimir energy-momentum tensors induced by a
fluctuating minimally coupled scalar field obeying Dirichlet boundary
conditions. We have demonstrated that this general problem can be
formulated with the aid of the worldline approach to the Casimir
effect. Contrary to simple interaction energy computations, the
worldline approach for such composite operators now has to deal with
open worldlines arising from the point-splitting procedure. As a proof
of principle, we have been able to show that the numerical worldline
approach to composite operators is able to reproduce analytically
known results for the single- and parallel-plate cases at low cost and
high efficiency. In particular, the single-plate case already shows
violations of the weak as well as the null energy condition. Of
course, the main goal of this project is to study more general classes
of Casimir configurations which particularly allow for non-trivial
tests of the averaged null energy condition. Work in this direction is
in progress. 

\noindent {\bf Acknowledgments:}
 We~have benefited from activities within the
ESF Research Network CASIMIR. This work has been supported by the DFG
under grant Gi328/5-1 (H.G.), Gi328/3-2 (I.H. and H.G.), GRK 1523 (M.S.) and partly by Conacyt (I.H.).

\end{document}